# Evidence for superconductivity above 260 K in lanthanum superhydride at megabar pressures

Maddury Somayazulu[1,*], Muhtar Ahart[1], Ajay K Mishra[2], Zachary M. Geballe[2], Maria Baldini[2], Yue Meng[3], Viktor V. Struzhkin[2], and Russell J. Hemley[1,*]

[1]*Institute for Materials Science and Department of Civil and Environmental Engineering, The George Washington University, Washington DC 20052, USA*
[2]*Geophysical Laboratory, Carnegie Institution of Washington, Washington DC 20015, USA*
[3]*HPCAT, X-ray Science Division, Argonne National Laboratory, Argonne IL 60439, USA*

Recent predictions and experimental observations of high $T_c$ superconductivity in hydrogen-rich materials at very high pressures are driving the search for superconductivity in the vicinity of room temperature. We have developed a novel preparation technique that is optimally suited for megabar pressure syntheses of superhydrides using pulsed laser heating while maintaining the integrity of sample-probe contacts for electrical transport measurements to 200 GPa. We detail the synthesis and characterization, including four-probe electrical transport measurements, of lanthanum superhydride samples that display a significant drop in resistivity on cooling beginning around 260 K and pressures of 190 GPa. Additional measurements on two additional samples synthesized the same way show resistance drops beginning as high as 280 K at these pressures. The drop in resistance at these high temperatures is not observed in control experiments on pure La as well as in partially transformed samples at these pressures, and x-ray diffraction as a function of temperature on the superhydride reveal no structural changes on cooling. We suggest that the resistance drop is a signature of the predicted superconductivity in $LaH_{10}$ near room temperature, in good agreement with density functional structure search and BCS theory calculations.

[*]zulu58@gwu.edu, rhemley@gwu.edu

The search for superconducting metallic hydrogen at very high pressures has long been viewed as a key problem in physics [1,2]. The prediction of very high (e.g., room temperature) $T_c$ superconductivity in hydrogen-rich materials [3] has opened new possibilities for realizing high critical temperatures but at experimentally accessible pressures (i.e., below 300 GPa) where samples can be characterized with currently available tools. Following the discovery of novel compound formation in the S-H system at modest pressures [4], theoretical calculations predicted that hydrogen sulfide would transform on further compression to a superconductor with a $T_c$ up to 200 K [5,6]. The high $T_c$ of 203 K at 150 GPa in samples formed by compression of $H_2S$ was subsequently confirmed [7,8], with x-ray measurements consistent with cubic $H_3S$ as the superconducting phase [9].

Given that the higher hydrogen content in many simple hydride materials is predicted to give still higher $T_c$ values [3], we have extended our studies to higher hydrides, the so-called superhydrides, $XH_n$ with $n > 6$. Systematic theoretical structure searching in the La-H and Y-H systems reveals numerous hydrogen-rich compounds with $T_c$ in the neighborhood of room temperature (above 270 K) [10] (see also Ref. [11]). Of these superhydrides, $LaH_{10}$ and $YH_{10}$ are predicted to have a novel clathrate-type structure with 32 hydrogen atoms surrounding each La or Y atom. These phases are predicted to exhibit strong electron-phonon coupling related to the phonon modes associated with the stretching of H-H bonds within the cages. Both are predicted to have superconducting temperatures near room temperature, i.e., near 270 K at 210 GPa for $LaH_{10}$ and 300 K at 250 GPa in $YH_{10}$. The superhydride phases consist of an atomic hydrogen sublattice with H-H distances of ~1.1 Å, which are close to the predicted values for solid atomic metallic hydrogen at these pressures [12]; for a thorough recent review of the field, see Ref. [13].

Recently, our group successfully synthesized a series of superhydrides in the La-H system up to 200 GPa pressures. Specifically, we reported x-ray diffraction and optical studies demonstrating that the lanthanum superhydrides can be synthesized with La atoms having an fcc lattice at 170 GPa upon heating to ~1000 K [14]. The heavy atom sublattice and equation of state are close to those of the predicted cubic metallic phase of $LaH_{10}$. Experimental and theoretical constraints on the hydrogen content give a stoichiometry of $LaH_{10\pm x}$, where $\pm x$ is between +2 and -1 [14]. On decompression, the fcc-based structure undergoes a rhombohedral distortion of the La sublattice, to form a structure that has subsequently been predicted to also have a high $T_c$ [12]. Here we report the use of a novel synthesis route for megabar pressure syntheses of such superhydrides using pulsed laser heating and ammonia borane as the hydrogen source. We detail the synthesis and characterization of several samples of the material using x-ray diffraction and electrical resistance measurements at 180-200 GPa. The transport measurements reveal a clear drop in resistivity on cooling at 260 K using four-probe measurements and as high as 280 K in other experiments. We infer that the resistance drop observed in these samples is a signature of nearly room-temperature superconductivity in $LaH_{10}$.

Samples were prepared for electrical conductivity measurements using a multi-step process with piston-cylinder diamond anvil cells [15]. Composite gaskets consisting of a tungsten outer annulus and a cubic boron nitride epoxy mixture (cBN) insert were employed to contain the sample at megabar pressures while isolating the electrical leads. Four platinum probes were sputtered onto the piston diamond (typically 1-2 $\mu$m thick) and connected to the external wires using a combination of 25-$\mu$m t☐☐☐☐☐platinum 'shoes' soldered onto brass holders [16]. The use of double beveled anvils loss of electrical contacts due to shearing of the intermediate connections was found to be less amenable to the loss of electrical contacts due to shearing of the intermediate connections at the gasket-diamond interface. Typically, we used 1/3 carat anvils with a 60-$\mu$m central culet beveled to 250 μm at 9° and to 450 $\mu$m at 8°.

While it is possible to load a mixture of La and $H_2$ into such a gasket assembly as in our previous work [14], we found that maintaining good contact between the synthesized sample and electrodes is not guaranteed. To overcome this problem, we used ammonia borane ($NH_3BH_3$, AB) as the hydrogen source. When completely dehydrogenated, one mole of AB yields three moles of $H_2$ plus insulating cBN, the latter serving both as a solid pressure medium and support holding the hydride sample firmly against the electrical contacts (Fig. 1). Although there is limited experimental data on the decomposition reaction kinetics of AB or related compounds above 50 GPa, our extensive earlier studies of the *P-T* dependence of dehydrogenation in AB (and related compounds by other groups) under pressure indicate both release of $H_2$, and no subsequent uptake of $H_2$ after release, at high pressures [17-19]. In addition, reports of catalytic dehydrogenation of AB presents the possibility of enhanced chemical activity at the La-AB interface [20].

In a typical run, a 5 μm thick sample of La was sandwiched between the electrodes and AB (typically 5-8 μm thick) in a cBN sample chamber about 40 μm in diameter. When pressure was increased, the rapid compression of the softer AB caused the cBN sample chamber to thin rapidly. We found that an initial cBN gasket thickness less than 15 $\mu$m guaranteed no implosion of the cBN gasket, which would in turn compromise the electrodes. AB was loaded in a glovebox in Ar environment with very low $O_2$ and $H_2O$ content (<0.1 ppm and <1 ppm, respectively). We also found that no chemical damage to the La sample occurred if it was loaded outside the glovebox within 30 mins of being removed, as this was needed to use a micromanipulator outside the glovebox to position the sample precisely on the piston diamond. The cell was clamped inside the glovebox and pressure typically raised to about 60 GPa.

After loading, synchrotron x-ray diffraction of the La-AB mixture was measured on the 16-ID-B beamline at Sector 16, HPCAT at the Advanced Photon Source, Argonne National Laboratory. Typically, a 5 x 5 $\mu$m focused x-ray beam at 0.4046 Å was rastered across the sample. This served as both the check on the chemical integrity of the sample as well as determining the spatial extent over which laser heating needed to be performed for synthesis. A typical data set showed no evidence for reactions taking place during rapid compression without heating. Further, no appreciable peak broadening occurred on compression, indicating that AB served as a reasonably quasihydrostatic pressure medium. Pressure was determined using x-ray diffraction from a gold pressure marker [21] and from the Raman shift of the diamond anvil tip with 660 nm and 532 nm laser sources [22]. The observed compression of the La sample determined by x-ray diffraction was found to coincide with a control run using Ne as a pressure medium [14], further confirming that the AB remained quasihydrostatic over these conditions.

These synthesis experiments were made possible by a versatile laser heating system [23] that was adapted for our experiments. *In-situ* laser heating of the sample with a Nd-YLF laser was performed at pressures above 185 GPa, since this was observed to be the lowest pressure for which $LaH_{10\pm x}$ is both synthesized and remains stable [14]. Only single-sided laser heating was performed since the synthesis proceeded from the La-AB interface, and this was the side that is buffered from both the diamond and the electrodes (Fig. 1). Pulsed laser heating was used in all our synthesis experiments to reduce damage to the anvils as well as to minimize formation of other reaction products mentioned below. In all runs, we observed very good coupling and subsequent sharpening of diffraction peaks followed by partial transformation of the sample to cubic $LaH_{10\pm x}$ when temperatures were above 1200 K. Above 2000 K, we observed additional diffraction peaks, including those indicating a monocline unit cell similar to that reported for Mg(BH$_2$)$_2$ [24]. To complete the transformation, we needed to maintain sample temperatures below 1800 K. This was achieved by varying the combination of laser power and pulse width of the heating laser. Typically, we found that a 300-ms pulse resulted in better transformation, better coverage and the lowest damage threshold. The laser spot size was kept to roughly 30 $\mu$m to cover the sample while chilled water was circulated around the cell to minimize movement due to thermal expansion of assembly components. This aspect of the system was crucial since movement of the cell would result in changes in laser alignment and/or laser spot size that could damage the electrodes. The sample was then rastered in the x-ray beam following the laser heating to confirm complete transformation. Typically, we found that the region in contact with the electrodes could be partially transformed [16], which is plausible since the electrodes would act as an additional heat sink with which the sample is in thermal contact. We also performed a check on the two-probe resistance between the four electrodes between heating cycles to ensure that the electrodes were not damaged.

Representative diffraction patterns of three samples are shown in Fig. 2, including a LaH$_{10\pm x}$ sample that was verified to have the four-probe geometry intact at 188 GPa. The characteristic diffraction peaks of LaH$_{10\pm x}$ identified previously from the synthesis using La and H$_2$ are observed, demonstrating that the superhydride phase can indeed be synthesized from La and AB using the techniques described above. Further, the results show that synthesis can be carried out while maintaining the electrical leads intact (Fig. 1). After synthesis, the cells were introduced into a flow-type cryostat (i.e., off-line from the x-ray measurements), and the leads connected to a function generator and lock-in amplifier for the electrical conductivity measurements [25]. Pressures were monitored from diamond Raman measurements after each cooling and warming cycle. While every effort was made to minimize pressure excursions during cooling cycles (by using opposing load screws), we found that in cells made of stainless steel, pressure increased during the first cooling cycle by as much as 10 GPa.

Figure 3 shows electrical resistance measurements as a function of temperature for the sample depicted in Fig. 1 at an initial 300 K pressure of 188 GPa. On cooling, the resistance was observed to decrease around 275 K and to drop appreciably at 260 K, as first reported in Ref.

[26]. The resistance abruptly dropped by >$10^3$ and remained constant from 253 K to 150 K. Upon warming, the resistance increased steeply at 245 K, indicating the change was reversible shifted to lower temperature. Upon subsequent warming to 300 K, the pressure measured to be 196 GPa. Since the pressure was not measured as a function of temperature during thermal cycling, the pressure at which the resistance change occurred was not measured. It is normal for the pressure to increase when cooling with the long piston-cylinder cells used in this experiment. Additional experiments were carried out to confirm these measurements. A summary of the experiments reported here, including the synthesis volumes, is provided in Fig. 4. Figures S2 and S3 show the results of additional resistance measurements performed with a pseudo-four probe configuration. A similar dramatic loss of resistance was observed, beginning between 250 but in one cycling extending up to 280 K at comparable pressures [16]. Together with the four-probe measurement data (Fig. 2) the results show abrupt changes in resistance between 245 K and 280 K in three samples on cooling and warming.

We further explored whether the observed resistivity transition at these high temperatures is associated with superconductivity. Additional experiments were carried out to determine whether the resistance change is due to a temperature-induced structural transition. To address this, and to identify other possible structural changes in the samples on thermal cycling, *in situ* x-ray diffraction was measured as a function of temperature after synthesizing $LaH_{10\pm x}$ from La and AB as indicated above. The x-ray diffraction patterns reveal no structural changes in the sample (Figs. S3 and S4). These measurements also indicated pressure changes $\Delta P < 0.1$ GPa on cycling from 300 K to 150 K. An alternative explanation for the change is a temperature-induced, iso-structural electronic transition with a dramatic increase in conductivity in the low-temperature phase not associated with superconductivity. Consistent with our previous optical observations [14], the resistance measurements indicate that the high-temperature phase (i.e., normal state) is a metal, so this would be metal-metal transition. However, calculations reported to date do not predict such a transition within metallic $LaH_{10}$ [10,12].

We did not attempt to determine the intrinsic resistivity of the superhydride samples because of the complex geometry and we recognize that the material could be mixed phase, possibly with varying hydrogen stoichiometry [10]. The samples could consist of layers of $LaH_x$ starting from $x = 10$ on the laser heated side and x <10 toward the electrodes. The samples could also have a complicated toroidal geometry consisting of a central region of LaH$_{10\pm x}$ with a peripheral ring of untransformed La arising from the need to preserve the electrodes during laser heating and differential thermal heat transport for the sample in contact with the diamond versus that in contact with the electrodes.

If the material is in fact superconducting, it is also possible that the variability in the transition temperature signifies a changing network of the superconducting component on thermal cycling. The increase in pressure affected the contact resistance as evidenced by a large decrease after the cooling cycle especially in samples that were probed with pseudo-four probe techniques. Changes in contact resistance would affect pseudo-four probe measurements, and this could in turn be affected by temperature-induced changes in the complex stress field of these samples (see 3 and Ref. [16]). The drop in resistance in the pseudo-four probe measurements (i.e., ~10%) is comparable to what has been observed in previous high-pressure resistance measurements [27,28]. The lowest measured resistance in the four-probe geometry was of order 20 $\mu\Omega$ (Fig. 3). In the three-point (pseudo-four probe) geometry, the contact resistance plays a major role and such low resistance values cannot be measured [27,29]. Although the complexity of the experiments prevents us from accurately determining the pressure dependence of the possible superconducting $T_c$, the four-probe measurements do suggest a decrease in the transition temperature with pressure, in agreement with theory [10].

In summary, we report resistance measurements on LaH$_{10\pm x}$ synthesized at pressures of 180-200 GPa by a pulsed laser technique that preserves the integrity of multi-probe electrical contacts on the sample after synthesis. These experiments remain challenging due to the requirement that the electrical probes not be damaged, and the limited temperature range of the synthesis. Nevertheless, our multiple measurements reveal the signature of a major resistivity transition at consistent high temperatures of 245-280 K at 190-200 GPa. The drop in resistance occurs around the same temperature as the superconducting $T_c$ predicted from BCS calculations for the LaH$_{10}$ structure at comparable pressures. Measurements of diamagnetic effects as well as direct measurements of the Meissner effect are crucial as well as understanding the dependence of $T_c$ on stoichiometry. Additional measurements along these lines are in progress, including additional constraints from optical and infrared spectroscopy. If confirmed, the measurements reported here provide the first experimental evidence of conventional superconductivity near room temperature in a new class of materials.

## Acknowledgements

We are grateful to H. Liu, S. Sinogeikin, I. I. Naumov, R. Hoffmann, N. W. Ashcroft, and S. A. Gramsch for their help in many aspects of this work. The authors would like to acknowledge the support of Paul Goldey and honor his memory. This research was supported by EFree, an Energy Frontier Research Center funded by the U.S. Department of Energy (DOE), Office of Science, Office of Basic Energy Sciences (BES), under Award DE-SC0001057. The instrumentation and facilities used were supported by DOE/BES (Award DE-FG02-99ER45775, VVS), the U.S. DOE/National Nuclear Security Administration (Award DE-NA-0002006, CDAC; and Award DE-NA0001974, HPCAT). The Advanced Photon Source is operated for the DOE Office of Science by Argonne National Laboratory under Contract DE-AC02-06CH11357.

# References

[1] N. W. Ashcroft, Metallic hydrogen - A high-temperature superconductor, *Phys. Rev. Lett.* **21**, 1748 (1968).

[2] V. L. Ginzburg, What problems of physics and astrophysics seem now to be especially important and interesting (thirty years later, already on the verge of XXI century)? , *Phys. Uspekhi* **42**, 353 (1999).

[3] N. W. Ashcroft, Hydrogen dominant metallic alloys: high temperature superconductors?, *Phys. Rev. Lett.* **92**, 187002 (2004).

[4] T. A. Strobel, P. Ganesh, M. Somayazulu, P. R. C. Kent and R. J. Hemley, Novel cooperative interactions and structural ordering in $H_2S$-$H_2$, *Phys. Rev. Lett.* **107**, 255503 (2011).

[5] Y. Li, J. Hao, H. Liu, Y. Li and Y. Ma, The metallization and superconductivity of dense hydrogen sulfide, *J. Chem. Phys* **140**, 174712 (2014).

[6] D. Duan, *et al.*, Pressure-induced metallization of dense $(H_2S)_2H_2$ with high-$T_c$ superconductivity, *Sci. Rep.* **4**, 6968 (2014).

[7] A. P. Drozdov, M. I. Eremets and I. A. Troyan, Conventional superconductivity at 190 K at high pressures, *arXiv:1412.0460*

[8] A. P. Drozdov, M. I. Eremets, I. A. Troyan, V. Ksenofontov and S. I. Shylin, Conventional superconductivity at 203 K at high pressures, *Nature* **525**, 73-76 (2015).

[9] M. Einaga, *et al.*, Crystal structure of the superconducting phase of sulfur hydride, *Nature Phys.* **12**, 835-838 (2016).

[10] H. Liu, I. I. Naumov, R. Hoffmann, N. W. Ashcroft and R. J. Hemley, Potential high-$T_c$ superconducting lanthanum and yttrium hydrides at high pressure, *Proc. Natl. Acad. Sci. USA* **114**, 6990-6995 (2017).

[11] F. Peng, Y. Sun, C. J. Pickard, R. J. Needs, Q. Wu and Y. Ma, Hydrogen clathrate structures in rare earth hydrides at high pressures: Possible route to room-temperature superconductivity, *Phys. Rev. Lett.* **119**, 107001 (2017).

[12] H. Liu, I. I. Naumov, Z. M. Geballe, M. Somayazulu, J. S. Tse and R. J. Hemley, Dynamics and superconductivity in compressed lanthanum superhydride, *submitted*

[13] T. Bi, N. Zarifi, T. Terpstra and E. Zurek, The search for superconductivity in high pressure hydrides, *arXiv:1806.00163*

[14] Z. M. Geballe, *et al.*, Synthesis and stability of lanthanum superhydrides, *Angew. Chem. Inter. Ed.* **57**, 688-692 (2018).

[15] H. K. Mao, R. J. Hemley and A. L. Mao, Recent design of ultrahigh-pressure diamond cell, in *High Pressure Science and Technology --1993*, edited by S. C. Schmidt, *et al.* (AIP Press, New York, 1994), pp. 1613-1616.

[16] *Supplementary Materials*

[17] R. S. Chellappa, M. Somayazulu, V. V. Struzhkin, T. Autrey and R. J. Hemley, Pressure-induced complexation of $NH_3BH_3$–$H_2$, *J. Chem. Phys.* 224515 (2009).

[18] Y. Song, New perspectives on potential hydrogen storage materials using high pressure, *Phys. Chem. Chem. Phys.* **15**, 14524–14547 (2013).

[19] R. G. Potter, M. Somayazulu, G. D. Cody and R. J. Hemley, High pressure equilibria of dimethylamine borane, dihydridobis(dimethylamine)boron(III) tetrahydridoborate(III), and hydrogen *J. Phys. Chem. C* **118**, 7280-7287 (2014).

[20] W.-W. Zhan, Q.-L. Zhu and Q. Xu, Dehydrogenation of ammonia borane by metal nanoparticle catalysts, *ACS Catalysis* **6**, 6892-6905 (2016).

[21] D. L. Heinz and R. Jeanloz, The equation of state of the gold calibration standard, *J. Appl. Phys.* **55**, 885 (1984).

[22] Y. Akahama and H. Kawamura, Pressure calibration of diamond anvil Raman gauge to 310 GPa, *J. App. Phys.* **100**, 043516 (2006).

[23] Y. Meng, G. Shen and H. K. Mao, Double-sided laser heating system at HPCAT for in situ x-ray diffraction at high pressures and high temperatures, *J. Phys. Cond. Matter* **18**, S1097 (2007).

[24] R. Černý, Y. Filinchuk, H. Hagemann and K. Yvon, Magnesium borohydride: synthesis and crystal structure, *Angew. Chem. Inter. Ed.* **46**, 5765-5767

[25] L. J. van der Pauw, A method of measuring specific resistivity and Hall effect of discs of arbitrary shape, *Philips Res. Rep.* **13**, 1-9 (1958).

[26] R. J. Hemley, Progress on hydride, superhydride, and hydrogen superconductors, *International Symposium: Pressure and Superconductivity. Fundacion Ramon Areces - Madrid, Spain, May 21-22* (2018).

[27] V. V. Struzhkin, M. I. Eremets, W. Gan, H. K. Mao and R. J. Hemley, Superconductivity in dense lithium, *Science* **298**, 1213-1215 (2002).

[28] R. Dias*, et al.*, Superconductivity in high disordered dense carbon disulfide, *Proc. Nat. Acad. Sci.* **110**, 11720-11724 (2013).

[29] A. M. Schaeffer, S. R. Temple, J. K. Bishop and S. Deemyad, High-pressure superconducting phase diagram of $^6$Li: Isotope effects in dense lithium, *Proc. Nat. Acad. Sci.* **112**, 60-64 (2015).

[30] A. Dewaele, P. Loubeyre and M. Mezouar, Equations of state of six metals above 94 GPa, *Phys. Rev. B* **70**, 094112 (2004).

[31] P. Loubeyre*, et al.*, X-ray diffraction and equation of state of hydrogen at megabar pressure, *Nature* **383**, 702-704 (1996).

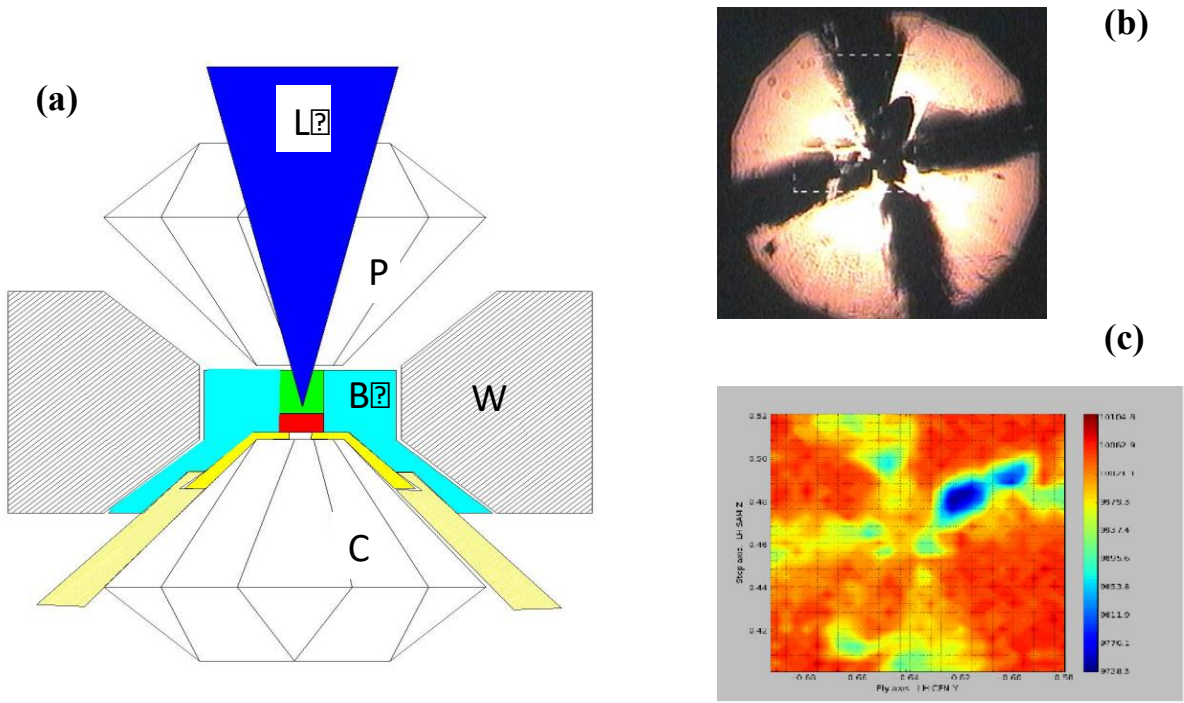

**Figure 1.** (a) Schematic of the assembly used for synthesis and subsequent conductivity measurements. The sample chamber consists of a tungsten outer gasket (W) with an insulating cBN insert (B). The piston diamond (P) is coated with four 1-$\mu$m thick Pt electrodes which are pressure bonded to 25-$\mu$m thick Pt electrodes (yellow). The 5-$\mu$m thick La sample (red) is placed on the Pt electrodes and packed in with ammonia borane (AB, green). Once the synthesis pressure is reached, the single-sided laser heating (L) is used to initiate the dissociation of AB and synthesis of $LaH_{10\pm x}$. To achieve optimal packing of AB without compromising the gasket strength, we loaded AB first with the gasket fixed on the cylinder diamond (C). (b) Optical micrograph of the sample at 178 GPa after laser heating. (c) X-ray transmission radiograph of the assembly in a similar orientation, which shows the sample confined to the junction of the four Pt electrodes. The radiograph was reconstructed from contouring the transmitted x-ray intensity ($\lambda$ = 0.4066 Å, or $E$ = 30.492 keV) on a photodiode behind the sample that was rastered in perpendicular x and y directions in 2-$\mu$m steps.

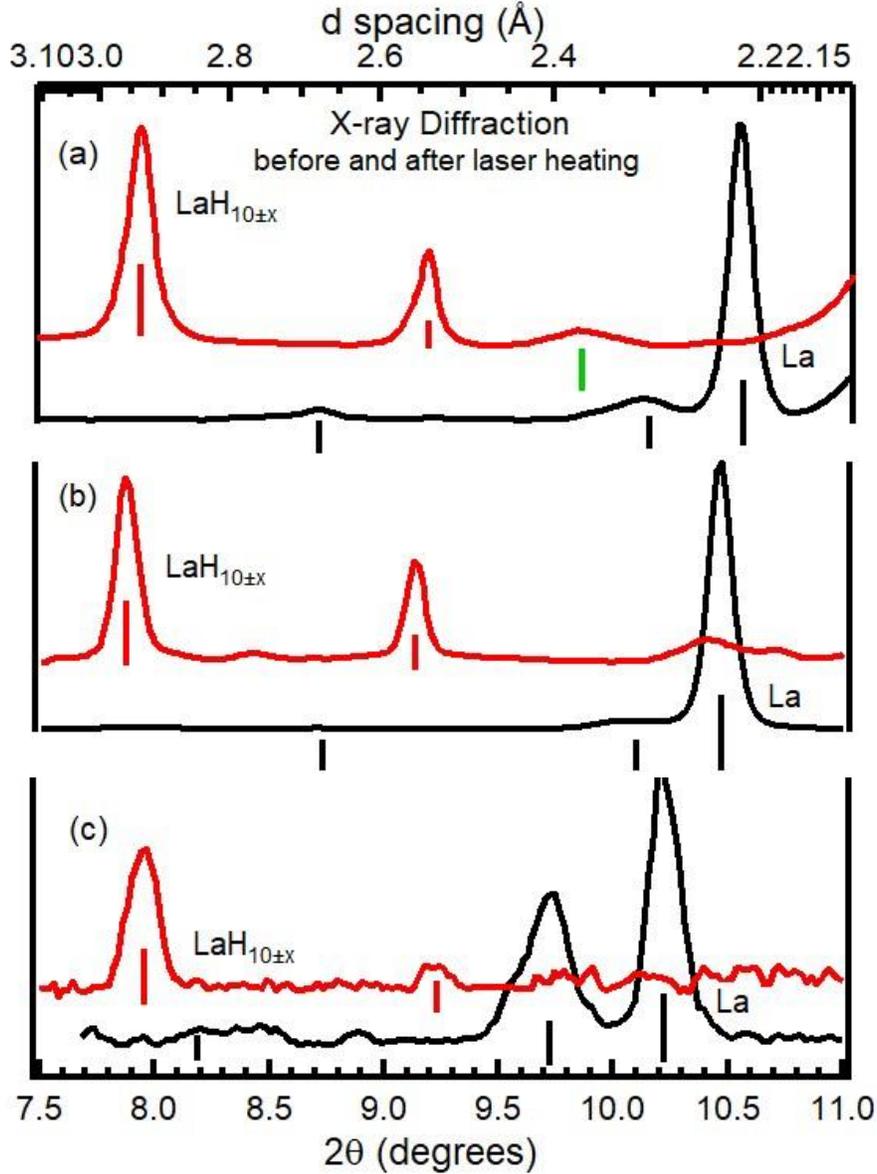

**Figure 2.** Synchrotron x-ray diffraction patterns obtained from three samples in which $LaH_{10\pm x}$ was synthesized by laser heating at pressures above 175 GPa. (a) X-ray diffraction following laser heating of a mixture of La and $H_2$; (b) and (c) results obtained following laser heating of La and $NH_3BH_3$. The data were obtained from three separate runs with a nominal x-ray spot size of 5x5 $\mu$m. In all three panels, the data shown in black are from the $hR24$ phase of La, and the pattern shown in red is from the transformed fcc-based $LaH_{10\pm x}$ obtained after laser heating above 175 GPa. The sample shown in (c) was nominally 5-$\mu$m thick and shows a high degree of stress (both in the relative intensities and the broadening of the diffraction peaks of the $hR24$ phase). The diffraction peak marked by the green symbol in (a) is residual $WH_x$, which is formed when $H_2$ is present. This peak is absent in the other two patterns obtained with the insulating cBN gasket.

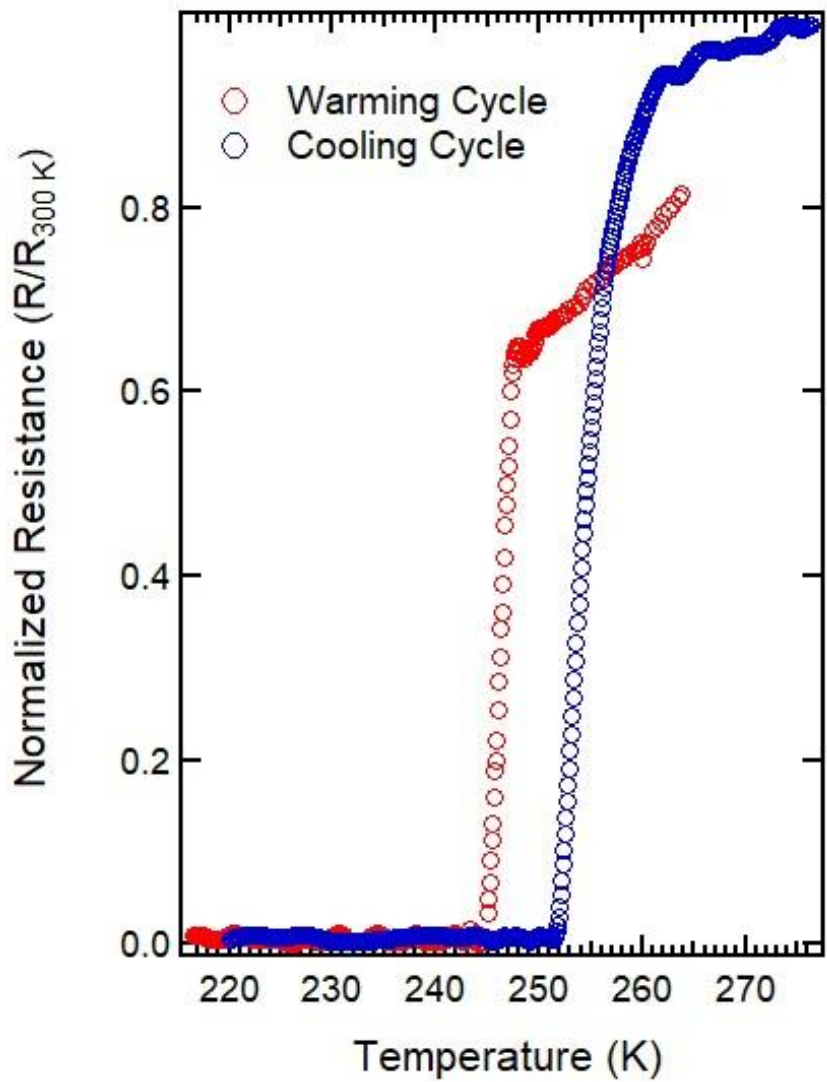

**Figure 3**. Normalized resistance of the $LaH_{10\pm x}$ sample characterized by x-ray diffraction and radiography (Figs. 1 and 2) and measured with a four-probe technique. The initial pressure was 188 GPa as determined from the Raman measurements of the anvil diamond edge; the pressure after the first cooling (blue) and warming (red) cycle was found to be 196 GPa. The lowest resistance we could record was 20 μΩ after the drop shown in the plot; the 300 K value was 50 mΩ. The measurements were performed using a EG&G model 5209 lock-in amplifier with nominally 10 mA current at 1 kHz with a typical cooling and warming rate of 1 K/min.

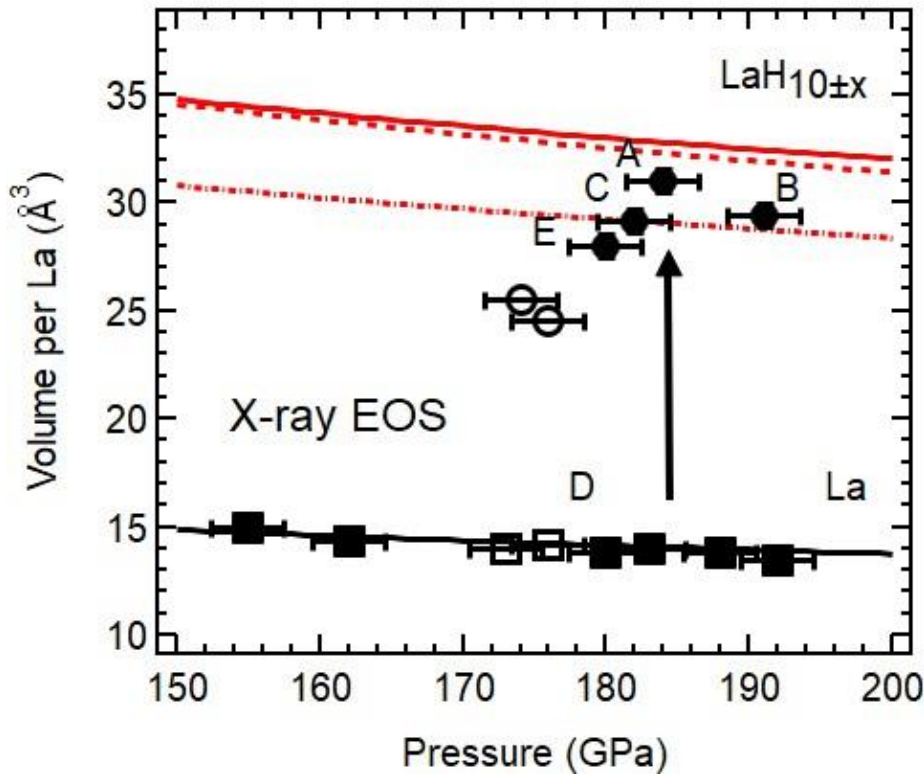

**Figure 4**. Observed volumes per La before and after the conversion from La to LaH$_{10\pm x}$ are shown for several samples for which low temperature conductivity measurements have been performed. Several other samples that lost electrical contacts have been excluded. All these samples used NH$_3$BH$_3$ as the hydrogen source and were laser heated at pressures higher than 170 GPa as determined by a combination of diamond Raman [22] and Pt equation of state [30]. The x-ray determined *P-V* equations of state we reported earlier for LaH$_{10\pm x}$ and La are shown by the solid lines [14]. The predicted *P-V* curves for the assemblages La + 5H$_2$ and La + 4H$_2$ are plotted as the dashed and dash-dot lines, respectively, using the equations of state for La [14] and H$_2$ [31]. The arrow indicates the superhydride synthesis, which is accompanied by conversion of the rhombohedral (*R*-3m) *hR*24 phase of La to cubic (*Fm*-3*m*) LaH$_{10\pm x}$ and an increase in the volume per La [14]. The two samples denoted by open symbols displayed no resistivity transition. Their volumes lie below the La + 4H$_2$ line, suggesting a threshold stoichiometry of LaH$_8$ (i.e., $x = -2$) for the transition. Further details are provided in Ref. [15].

# Evidence for superconductivity above 260 K in lanthanum superhydride at megabar pressures

Maddury Somayazulu[1,*], Muhtar Ahart[1], Ajay K Mishra[2], Zachary M. Geballe[2], Maria Baldini[2], Yue Meng[3], Viktor V. Struzhkin[2], and Russell J. Hemley[1,*]

[1]*Institute for Materials Science and Department of Civil and Environmental Engineering, The George Washington University, Washington DC 20052, USA*
[2]*Geophysical Laboratory, Carnegie Institution of Washington, Washington DC 20015, USA*
[3]*HPCAT, X-ray Science Division, Argonne National Laboratory, Argonne IL 60439, USA*

*zulu58@gwu.edu, rhemley@gwu.edu

We have developed a novel preparation technique that is optimally suited for megabar pressure syntheses of superhydrides using pulsed laser heating while maintaining the integrity of sample-probe contacts for electrical transport measurements to 200 GPa. We report the synthesis and characterization, including four-probe electrical transport measurements, of lanthanum superhydride samples that display a significant drop in resistivity on cooling beginning around 260 K and pressures of 190 GPa in our main text. In this supplementary material, we include some of the details on pressure measurements, analyses of low temperature x-ray diffraction, and analyses of electrical conductivity data.

Samples were prepared using a multi-step process with various diamond anvil cells [1,2]. Compound gaskets consisting of a tungsten outer annulus and a cubic boron nitride (cBN) insert were employed to encapsulate the sample at megabar pressures while isolating the electrical leads. Samples synthesized using *in-situ* diffraction and laser heating were then loaded into a cryostat for conductivity measurements.

The samples used for the electrical conductivity measurements and in low temperature x-ray diffraction are shown in Table S1. The synthesis pressures listed are the pressures before laser heating, based on the diamond Raman measurements [3], and corroborated with Pt and La equations of state [4,5]. The cell volumes listed are the volume per formula unit and are based on the position of the (111) diffraction peak of LaH$_{10\pm x}$ (*Fm-3m*). The (110) and (024) diffraction peaks of La were used to determine the cell volume of its *hr*24 (*R-3m*) phase.

Depending on the peak temperature, the heating laser pulse width, the thickness of the La sample, and the gasket thickness (which dictates the La to NH$_3$BH$_3$ ratio), we synthesize cubic LaH$_{10\pm x}$ (Fig. 3). In our previous work, we constrained $\pm x$ to be between +2 and -1 [5]. On cooling, samples A-C all exhibit abrupt resistance drops beginning as high as 280 K, whereas the resistance in sample D (untransformed La) monotonically decreases, characteristic of a normal metal. The resistance in the two samples denoted by open symbols were also measured but displayed no abrupt resistance drops on cooling, despite the fact that the four probes for both remained intact after synthesis. The diffraction data thus indicate that samples A, B, C, and E have substoichiometric hydrogen but maintain a cubic La structure. The samples that did not show evidence of the resistance drop between 300 and 80 K display a cubic symmetry but their volumes lie below the La + 4H$_2$ line, suggesting a threshold stoichiometry of LaH$_8$ (i.e., $x = -2$) for the material to exhibit the resistance drops documented here. The synthesis conditions of the samples for which data are reported here are listed in Table S1.

The measurements on LaH$_{10\pm x}$ were made using a three-point (pseudo four-probe) geometry and therefore the observed resistance is a sum of the sample resistance and the contact resistance [6] (Fig. S1). The normalized resistance R(T)-R$_c$(T)/R(300 K)-R$_c$(T), where R$_c$ is a linear fit to resistance at temperatures below the transition extrapolated to temperatures above the transition. The estimated contact resistance as a function of temperature R$_{300}$ is the ambient temperature resistance of LaH$_{10\pm x}$ along with that of pure La measured with the four-probe technique are shown in the same figure. Although the apparent increase in $T_c$ with pressure increase contradicts both theory [7] and the results of the four-probe measurement shown in Fig. 3, the behavior could represent changes in sample connectivity and stoichiometry (see main text) similar to that reported in H$_3$S [8].

Figure S2 shows an example of steps used in correcting for the contact resistance and obtaining the normalized sample resistance as a function of temperature. As is evident from the second panel, this ratio drops at about 250 K and shows a linear decrease thereafter. We assume that the linear resistance below this drop signifies a residual contact resistance which is subtracted out of the whole curve to produce the curve in the lowest panel. This would therefore correspond nominally to a relative resistance corrected for a contact resistance. The same procedure was applied to several other measurements made on this sample (B) as shown in Fig. S1.

The results of the low-temperature x-ray diffraction results are shown in Fig. S3 and S4. Due to the large 'footprint' of the cryostat (a flow type L-$N_2$ cryostat equipped with sapphire windows), a beam defining, collimating pinhole could not be placed close to the sample and correspondingly, a larger incident beam interrogated several parts of the sample at the same time. Further, due to differential contraction of several components in the cryostat, it is very difficult to interrogate the exact same part of the sample at different temperatures. We therefore obtained several diffraction patterns across a 50 x 50 $\mu$m grid (in 10-$\mu$m steps) and analyzed all of them. The data were fitted with several diffraction peaks comprising the sample, the pressure marker and cryostat window. This was a preferred method for extracting the d-spacings of Pt and $LaH_{10\pm x}$ since any profile refinement would not be able to account for pressure distribution as well as sample inhomogeneity. Apart from the stronger diffraction peaks ascribed to the cryostat exit window, we see no evidence for any new diffraction that would indicate a structural transition in cubic $LaH_{10\pm x}$ in the 80 – 300 K temperature interval at 185 GPa. The lines show that there is neither a large pressure drift nor change in the *P-V* dependence of $LaH_{10\pm x}$ in this temperature range. In general, our experience with the stainless steel (SS) long piston-cylinder cells equipped with SS spring washers has been that the pressure increases on cooling (at least in the first cycle) by about 10-20 GPa due to changes in washer dimensions and thereby the load. In this experiment, we used non-magnetic cells and Be-Cu spring washers [2], which show smaller pressure drifts due to changes in load.

**Table S1**: Samples used for electrical conductivity and low-temperature x-ray diffraction measurements.

| Sample | Synthesis Pressure | Cell Volume | Experimental Details |
|---|---|---|---|
| **A** | 185(±5) GPa | 32.0(2) Å$^3$ | Four probe electrical conductivity |
| **B** | 188(±5) GPa | 31.1(2) Å$^3$ | Pseudo-four probe conductivity |
| **C** | 175(±5) GPa | 29.8(2) Å$^3$ | Pseudo-four probe conductivity |
| **D** | 180(±5) GPa | 14.5(2) Å$^3$ | Four probe electrical conductivity |
| **E** | 182(±5) GPa | 27.8(2) Å$^3$ | Low temperature x-ray diffraction |

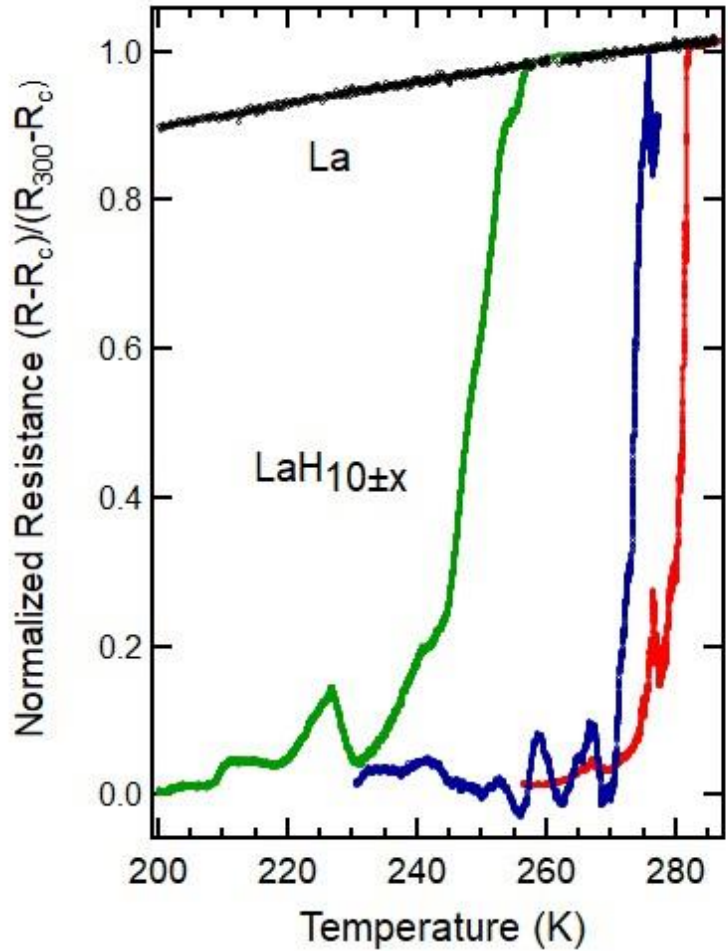

**Figure S1.** The normalized resistance $(R(T)-R_c(T))/(R(300\ K)-R_c(T))$ of $LaH_{10\pm x}$ (sample B) along with that of pure La (sample D) measured with a pseudo-four probe technique. The pressure measured at room temperature with the diamond Raman gauge [3] increased gradually with every cooling and warming cycle, as determined from measurements performed after the sample was extracted from the cryostat; i.e., 190 GPa (green), 195 GPa (blue) and 202 GPa (red) for the three warming and cooling cycles. A correction for a linear contact resistance was applied since this was a three-point measurement.

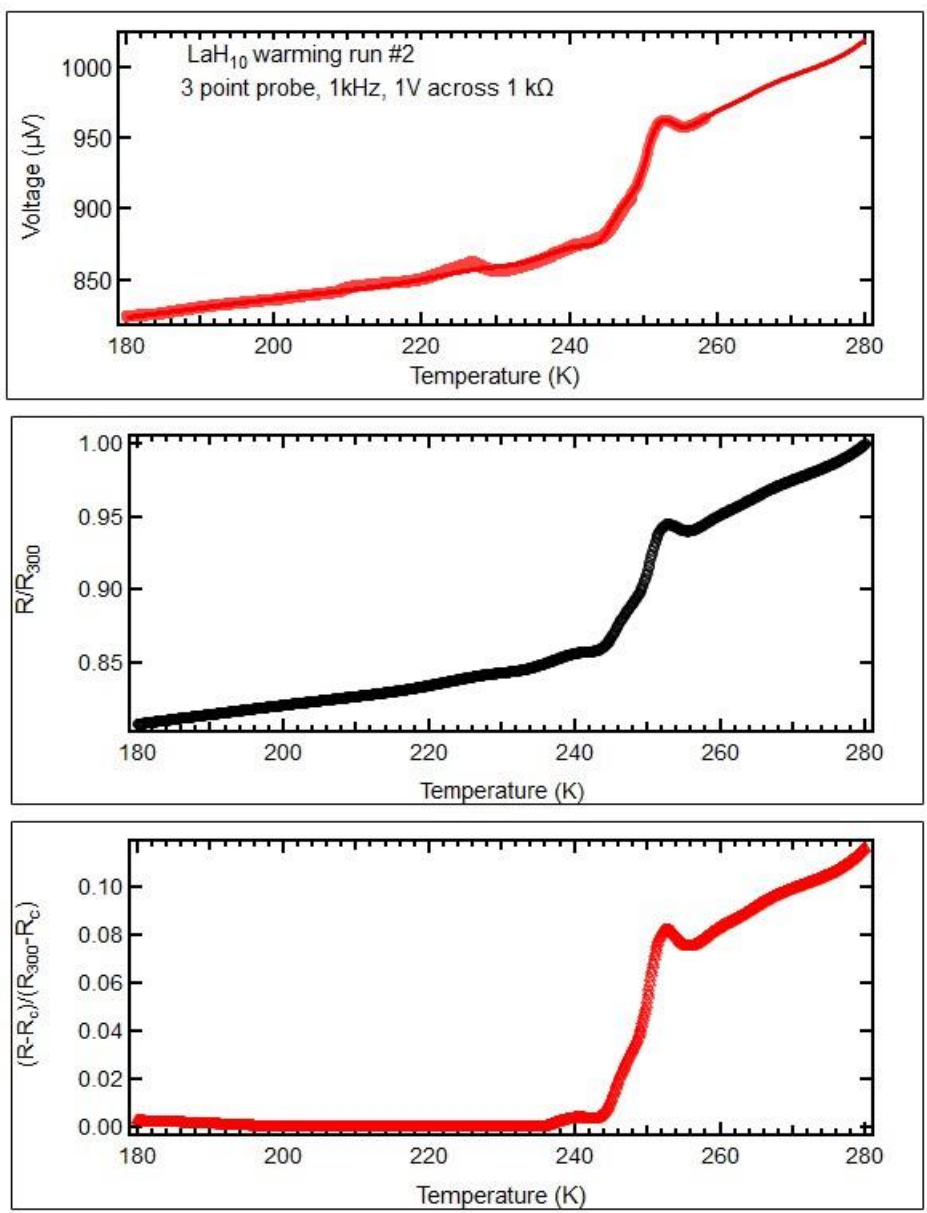

**Figure S2.** Selected resistance data obtained from the pseudo-four probe measurement of LaH$_{10\pm x}$ during a warming cycle from 150 K. The raw data shown in the top panel was normalized to the ambient resistance value (in this case, an extrapolated value) to obtain the relative resistance (R/R$_{300}$). These measurements were all made using a 1 V (peak-peak) sine wave input from a Rigol function generator applied to the sample using a 1 kΩ resistance to limit the load current as well as stabilize the applied load current. The output voltage was measured in a differential mode on a EG and G model 5209 lock-in amplifier with 1 kΩ isolating resistances across the input to the common ground.

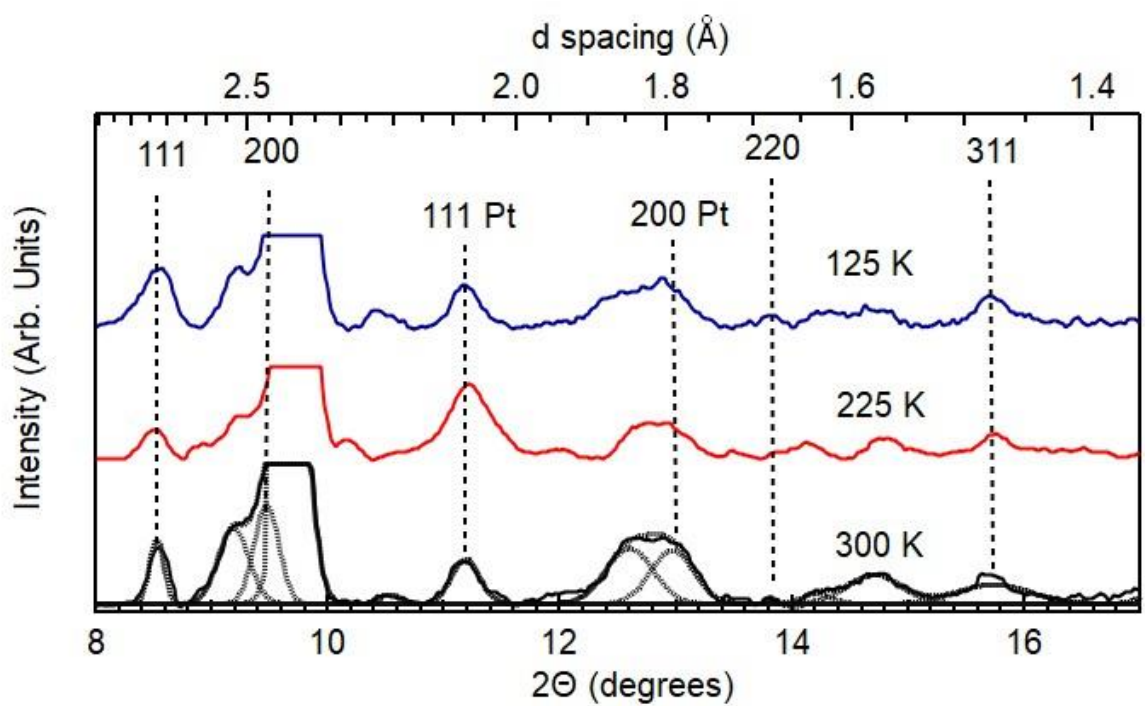

**Figure S3:** Powder x-ray diffraction patterns obtained during cooling of LaH$_{10\pm x}$ (sample E) after synthesis at 182 GPa. The indices of the principal peaks of LaH$_{10\pm x}$ except where identified as Pt. The fit for the 300 K pattern is indicated, and only peaks for LaH$_{10\pm x}$ and Pt are shown. Additional peaks in all patterns arise from background scattering from the sapphire window of the cryostat. The position of 200 LaH$_{10\pm x}$ peak overlaps two background features and is therefore more uncertain as reflected in the larger error bars shown in Fig. S4.

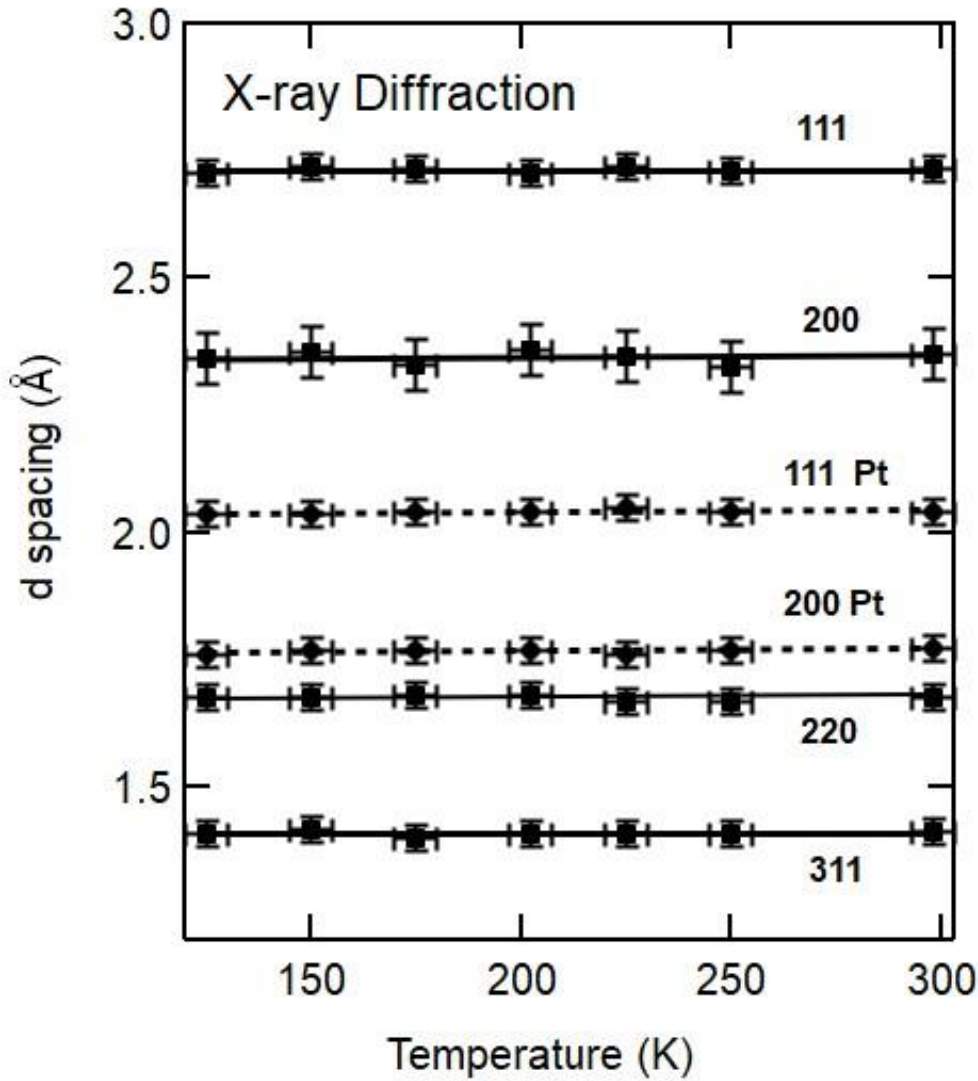

**Figure S4**: Observed d-spacings of LaH$_{10\pm x}$ and Pt as a function of temperature at 185 GPa taken from analysis of patterns such as those shown in Fig. S3. The error bars for the 200 line of LaH$_{10\pm x}$ are larger because they were extracted from a fit to three overlapping diffraction peaks (see Fig. S3). The parameters of the accompanying, stronger diffraction peaks arising from the cryostat windows were fixed in the refinement. Coupled with this is also the effect of pressure broadening which manifests itself in at least two diffraction peaks corresponding to two pressures that are offset by at least 20-30 GPa. We choose the highest-pressure peak (although it could be smaller in intensity) and display this position in the figure above with the peak width as the error bar. The error in pressure is determined from the width and indeterminacy of the Pt diffraction peaks.

# References

[1] H. K. Mao, R. J. Hemley and A. L. Mao, Recent design of ultrahigh-pressure diamond cell, in *High Pressure Science and Technology --1993*, edited by S. C. Schmidt, *et al.* (AIP Press, New York, 1994), pp. 1613-1616.

[2] A. G. Gavriliuk, A. A. Mironovich and V. V. Struzhkin, Miniature diamond anvil cell for broad range of high pressure measurements, *Rev. Sci. Instr.* **80**, 043906 (2009).

[3] Y. Akahama and H. Kawamura, High-pressure Raman spectroscopy of diamond anvils to 250 GPa: Method for pressure determination in the multimegabar pressure range, *J. Appl. Phys.* **96**, 3748-3751 (2004).

[4] A. Dewaele, P. Loubeyre and M. Mezouar, Equations of state of six metals above 94 GPa, *Phys. Rev. B* **70**, 094112 (2004).

[5] Z. M. Geballe, *et al.*, Synthesis and stability of lanthanum superhydrides, *Angew. Chem. Inter. Ed.* **57**, 688-692 (2018).

[6] M. I. Eremets and I. A. Troyan, Conductive dense hydrogen, *Nat. Mater.* **10**, 927-931 (2011).

[7] H. Liu, I. I. Naumov, R. Hoffmann, N. W. Ashcroft and R. J. Hemley, Potential high-$T_c$ superconducting lanthanum and yttrium hydrides at high pressure, *Proc. Natl. Acad. Sci. USA* **114**, 6990-6995 (2017).

[8] A. P. Drozdov, M. I. Eremets and I. A. Troyan, Conventional superconductivity at 190 K at high pressures, *arXiv:1412.0460*